\documentclass[useAMS]{mn2e}
\usepackage{times}
\usepackage{amssymb}
\usepackage{rotating}
\input{epsf}

\title[Frame dragging and the iron line]
{The effect of frame dragging on the iron K$_\alpha$ line in X-ray binaries}
\author[A. Ingram \& C. Done]
{Adam
Ingram$^{1}\thanks{E-mail:a.r.ingram@durham.ac.uk}$ \&
Chris Done$^{1}$ \\
$^1$Department of Physics, University of Durham, South Road,
Durham DH1 3LE, UK\\
}

\date{Submitted to MNRAS}
\pagerange{\pageref{firstpage}--\pageref{lastpage}} \pubyear{2012}

\begin{document}

\topmargin = -0.5cm

\maketitle

\label{firstpage}

\begin{abstract}

The clear characteristic timescale picked out by the low frequency
quasi-periodic oscillations (QPOs) seen in many black hole and neutron
star binaries has the potential to provide a very powerful diagnostic
of the inner regions of the accretion flow. However, this potential
cannot be realised without a quantitative model for the QPO.  We have
recently shown that the same truncated disc/hot inner flow geometry
which is used to interpret the spectral transitions can also directly
produce the QPO from Lense-Thirring (vertical) precession of the hot
inner flow. This correctly predicts both the frequency and spectrum of
the QPO, and the tight correlation of these properties with the total
spectrum of the source via a changing truncation radius between the
disc and hot flow. This model predicts a unique iron line signature as
a vertically tilted flow illuminates different azimuths of the disc as
it precesses. The iron line arising from this rotating illumination is
blue shifted when the flow irradiates the approaching region of the
spinning disc and red shifted when the flow irradiates the receding
region of the disc. This gives rise to a characteristic rocking of the
iron line on the QPO frequency which is a necessary (and probably
sufficient) test of a Lense-Thirring origin. This is also an
independent test of disc truncation models for the low/hard state, as
vertical precession cannot occur if there is a disc in the midplane.

We show that it may be possible to observe this effect using archival
data from the Rossi X-ray timing explorer (\textit{RXTE}) or
\textit{XMM Newton}. However, a clean test requires a combination of 
moderate resolution and good statistics, such as would be available
from a long XMM-Newton observation or with data from the proposed 
ESA mission \textit{LOFT}.

\end{abstract}


\section{Introduction} \label{sec:introduction}

Low frequency quasi-periodic oscillations (QPOs) are commonly observed
in the X-ray flux of both neutron star and black hole binaries (NSBs
and BHBs respectively; collectively X-ray binaries XRBs). They are
most clearly observed as strong, coherent features in the power
spectral density (PSD) which are Lorentzian in shape and so can be
described by amplitude (i.e. fractional rms variability), centroid
frequency ($f_{QPO}$) and width ($\Delta f$). These properties are
observed to be tightly correlated with the spectral properties of the
source which vary dramatically as the source rises from quiescence to
outburst before falling once more into quiescence (see e.g. van der
Klis 2006; Belloni 2010).

The physical processes behind this spectral evolution are
comparatively well understood. The spectral energy distribution (SED)
consists of three main components: a quasi-thermal disc, a power law (with
high and low energy cut-offs) and a reflection spectrum. When the source
flux is low (low/hard state), the power law is hard (photon index $\Gamma
\sim 1.7$) and dominates the SED. As the source flux increases, the power
law softens and weakens while the disc and reflection spectra increase in
luminosity (intermediate state). Eventually at the peak of the outburst,
the disc completely dominates (high/soft state), although sometimes there
is also a strong high energy tail (very high state). The disc spectrum is
well explained by a standard thin disc (Shakura \& Sunyaev 1973) and the
power law can be reproduced by Compton up-scattering of disc photons by
hot electrons in an optically thin corona. A fraction of the luminosity
emitted from the corona will then reflect off the disc to give a
reflection spectrum with the most obvious features being a strong iron
K$_\alpha$ line and a $\sim 30$ keV hump (see e.g. Fabian et al 2000). 

The truncated disc model, in which the thin disc only extends down to
some radius $r_o$, can naturally explain the evolution of the SED
(Esin et al 1997; Done, Gierlinski \& Kubota 2007). Interior to $r_o$
is a large scale height, optically thin accretion flow (hereafter, the
flow) which acts as the Comptonising corona. As the source flux
increases, the truncation radius moves inwards, thus increasing the
flux of disc photons incident on the flow and softening the power law
emission while simultaneously decreasing all characteristic timescales
associated with $r_o$. In this picture, $r_o$ moves from $\sim 60 -6$
(in units of $R_g=GM/c^2$) during the rise to outburst and back out
again during the fall back to quiescence.

QPOs are observed in the PSD during both the rise and the fall. During
the rise, the PSD displays a QPO (with harmonics) superimposed on a
broad band noise of variability.  The broad band noise can be roughly
characterised by two zero-centred Lorentzians with widths $f_b$ and
$f_h$ (e.g. Belloni, Psaltis \& van der Klis 2002). The QPO frequency
moves from $\sim 0.1-10$ Hz as the source flux increases and is
correlated with rises in both $f_b$ and $f_h$ (Wijnands \& van der
Klis 1999; Psaltis, Belloni \& van der Klis 1999). For BHBs, this is
commonly classified as the type-C QPO. Eventually the broad band noise
disappears and the PSD is dominated by a type-B QPO which has
$f_{QPO}\sim 6$ Hz. Before the source completely transitions into the
high/soft state, type-A QPOs are observed which are much broader and
weaker features, typically centred at $f_{QPO}\sim 8$ Hz. During the
fall, the same is observed in reverse (see Casella, Belloni \& Stella
2005 for more details of the A,B,C classification system and Belloni
2010 for a review of hysteresis behaviour).  Because these three types
of QPO are not observed simultaneously (even though the transition
between type-A and type-B QPOs can be very rapid) and they occupy a
similar frequency range, it is possible that they are driven by three
different variants of the same underlying physical process.

NSBs display a similar phenomenology of QPO types and spectral
transitions (although the nomenclature is very different; see e.g. van
der Klis 2005 for details). There is very strong evidence that the
QPOs in both classes of object are produced by the same process. The
correlations between $f_{QPO}$, $f_b$ and $f_h$ hold, with the same
gradient, for both NSBs and BHBs (Wijnands \& van der Klis 1999;
Klein-Wolt \& van der Klis 2008) with the only difference in frequency
being entirely consistent with mass scaling (Ingram \& Done 2011).

The physical process responsible for driving the QPO remains poorly
understood.  There are many QPO mechanisms suggested in the literature
(e.g. Fragile, Mathews \& Wilson 2001; Wagoner et al 2001; Titarchuk
\& Oscherovich 1999; Cabanac et al 2010; Tagger \& Pellat 1999;
O'Neill et al 2011; Kato 2008; Wang et al 2012).  However, such a rich
phenomenology means that these models are rarely able to explain all of
the known QPO properties simultaneously. In Ingram, Done \&
Fragile (2009), we suggested perhaps the most promising QPO model to
date. Based on the model of Stella \& Vietri (1998), we considered the
QPO to result from Lense-Thirring precession. This is a relativistic
effect which occurs because a spinning compact object drags spacetime
as it rotates. The orbit of a particle which is outside the plane of
black hole spin will therefore undergo precession because the starting
point of the orbit rotates around the compact object.  Stella \&
Vietri (1998) and Stella, Vietri \& Morsink (1999) showed that the
Lense-Thirring precession frequency of a test mass at the truncation
radius is broadly consistent with the QPO frequency. Schnittman (2005)
and Schnittman, Homan \& Miller (2006) developed this into a fully
relativistic description of a misaligned ring, showing that its direct
emission and iron line signature should be modulated on the precession
frequency, which could be somewhat higher than observed.
However, the real problem with these models is that the energy
spectrum of the QPO is dominated by the Comptonised emission
(Sobolewska \& Zycki 2006; Rodriguez et al 2004), requiring that the
QPO mechanism predominantly modulates the hot flow rather than the
disc (although the variability could be produced elsewhere before
propagating into the flow; Wilkinson 2011). We consider instead a
global precession of the entire hot flow, which naturally explains the
QPO spectrum. Such global precession has been seen in recent numerical
simulations (Fragile et al 2007, Fragile 2009).  We show in Ingram,
Done \& Fragile (2009) that the predicted frequency range is
completely consistent with the type-C QPO in BHBs and also in NSBs
(Ingram \& Done 2010).

There are other more subtle properties that are naturally predicted by
the precessing flow model. Heil, Vaughan \& Uttley (2011) show that
the QPO frequency is linearly related to the source flux on short
timescales ($\sim 3$s). We show in Ingram \& Done (2011) that
propagating fluctuations in mass accretion rate which give rise to the
broad band noise (e.g. Lybarskii 1997; Arevalo \& Uttley 2006) will
affect the moment of inertia of the flow leading the precession
frequency to fluctuate. The linear relation with flux then occurs
because both the flux \textit{and} the precession frequency depend on
mass accretion rate. Although it is very encouraging that this
property is predicted by the model, we still do not have unambiguous
proof that the flow precesses - a QPO produced from any mode of the
hot flow will also couple to fluctuations propagating through the hot
flow, and should give an $f_{QPO}$-flux relation.

The interpretation of the QPO as vertical precession requires a
truncated disc as otherwise the flow could not cross the equatorial
plane.  The issue of whether or not the disc truncates is still
somewhat controversial. The line clearly depends on
the spectral state, with a very small narrow line seen in the
dimmest low/hard states (e.g. Tomsick et al 2009), and a very broad line
when the source is very close to the transition to the soft state
(e.g. Hiemstra et al 2011; a hard intermediate state just after the 
transition from the soft state). However, in the brighter low/hard states,
Nowak et al (2011) show that the broad iron
line in  Cyg X-1 can be variously interpreted as
implying a disc anywhere from $6-32R_g$ (for their Obs 4) depending
whether the continuum is thermal Comptonisation, non-thermal
Comptonisation, multiple Compton components or includes a jet
contribution. Fabian et al (2012) show another deconvolution of a
similarly shaped spectrum from Cyg X-1, where the spectrum below
10~keV is dominated by highly ionised, highly smeared reflection, with
a very small inner radius of $\sim 1.3 R_g$ and a very steep
emissivity profile (a.k.a. the lightbending model). We note that this
lightbending geometry is inconsistent with the independent requirement
on the un-truncated disc geometry that the source is beamed away from
the disc in order to produce an intrinsically hard spectrum (Malzac,
Beloborodov \& Poutanen 2001). 

The issue is clearly still very controversial, though we note that the
rapid spectral variability can only currently be explained with an
inhomogeneous Comptonisation continuum model (Kotov et al 2001;
Arevalo \& Uttley 2006), where the line profile is consistent with a
truncated disc (Makishima et al 2008). Here we simply assume the
truncated disc geometry, and use this to propose a distinctive test of
a {\em vertical precession} origin of the QPO.  As a tilted flow
precesses, the illumination pattern on the disc rotates. The resulting
iron line is boosted and blue shifted at a time when the flow
illuminates the approaching side of the disc, and red shifted when the
flow illuminates the receding side of the disc. Since this periodic
rocking of the iron line is a requirement of the Lense-Thirring QPO
model, this also offers a potentially unambiguous test of disc
truncation. Our geometry differs from the Schnittman et al (2006)
model, where a precessing inner disc ring producing the iron line and
continuum. Instead, we have a hot inner flow replacing the inner disc
to produce the continuum, and precession of the entire hot flow
produces a rotating illumination pattern which excites the iron line
from the outer thin disc.

The paper is ordered as follows.  In section
\ref{sec:geometry}, we define the accretion geometry assumed for the
model. In section \ref{sec:model}, we will calculate the implications
of our assumed geometry on a very simple toy spectral model. In
section \ref{sec:spec}, we will take this further by introducing a
reasonable spectral model before analysing the likelihood of observing
this effect in section \ref{sec:obs}.

\section{Model geometry} \label{sec:geometry}

In this section, we outline the geometry used for our QPO model. We
assume that the spin axis of the compact object is misaligned with
that of the binary system as may be expected from supernova kicks
(Fragos et al 2010). Due to frame dragging, the orbit of an accreting
particle from the binary partner will precess around the spin axis of
the compact object. The effect of frame dragging on an entire
accretion flow depends on the dynamics of the flow. A thin accretion
disc being fed by a binary partner out of the spin plane of the
compact object will form a Bardeen Petterson configuration (Bardeen \&
Petterson 1975) where the outer regions align with the binary partner
and the inner regions align with the spin of the compact object, with
a transition between the two regimes at $r_{BP}$. The value of
$r_{BP}$ is not well known, with analytical estimates
ranging from $\sim 10 - 400~R_g$ (see e.g. Bardeen \& Petterson 1975;
Papaloizou \& Pringle 1983; Fragile, Mathews \& Wilson 2001).  In the
thin disc regime, warps caused by the misaligned black hole propagate
in a viscous manner. This means that the time scale on which a warp is
communicated is much longer than the precession period and therefore a
steady configuration forms. In contrast, warps in a large scale height
accretion flow are communicated by bending waves (see e.g. Lubow,
Ogilvie \& Pringle 2002; Fragile et al 2007) which propagate on
approximately the sound crossing timescale which is \textit{shorter}
than the precession period. For this reason, the hot flow can precess
as a solid body with the precession period given by a surface density
weighted average of the point particle precession period at each
radius (Liu \& Melia 2002), while a cool disc forms a stable warped
configuration. This solid body precession of a hot flow has been seen
explicitally in recent numerical simulations (Fragile et al 2007)
for the special case of a large scale height flow which we consider
here.

The key aspect is that the flow angular momentum has to be misaligned
with the black hole spin. Yet the outer thin disc will warp into
alignment with the black hole at $r_{BP}$. Since this radius is poorly
known, there are two possible scenarios.  Firstly $r_{BP}$ may be
small enough for the outer thin disc to still be aligned with the
binary partner at the truncation radius. In this case, the hot flow is
misaligned with the black hole spin by the intrinsic misalignment of
the binary system which will naturally lead to solid body precession of
the entire flow. Secondly, if $r_{BP}$ is large, the disc and hence hot
flow are intrinsically aligned with the black hole spin. However,
precession may be possible. The flow has a large scale height, so is
sub-Keplarian. At the truncation radius it overlaps with the Keplarian
disc, so this overlap layer is probably Kelvin-Helmholtz unstable,
producing turbulence. Clumps forming from random density fluctuations
in regions high above the midplane could temporarily mis-align the flow
leading to intermittent precession. This predicted intermitency has
the advantage of naturally explaining the observed random jumps in QPO
phase (Miller \& Homan 2005; Lachowicz \& Done 2010).

Here we assume the first geometry i.e.  assume that $r_{BP}$ is very
small. However, the effect of rotating illumination on the iron line
is qualitatively the same in the second geometry, differing only in
the details. In the next section, we outline the geometry used. We
work under the assumption that the central object is a black hole, 
but the geometry is valid for neutron stars also. 

\subsection{Disc}
\label{sec:disc}

\begin{figure}
\centering
\leavevmode  \epsfxsize=7.5cm \epsfbox{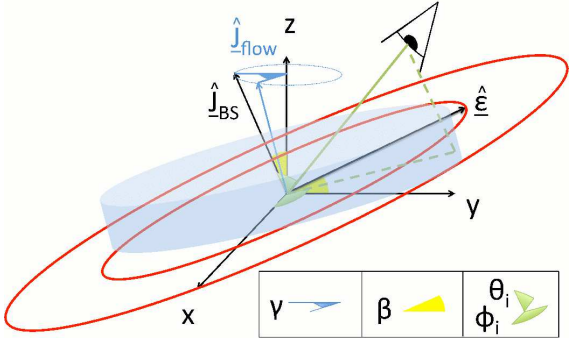}
\caption{Schematic diagram illustrating the coordinate system we are
considering. The black hole is at the origin and the black hole
angular momentum vector is aligned with the
z-axis. $\hat{\underline{J}}_{BS}$ is the (unit) angular momentum vector of
the binary system (and the disc), misaligned with the z-axis by
an angle $\beta$. $\hat{\underline{\epsilon}}$ then completes a right handed
Cartesian coordinate system
$\{x,\hat{\underline{\epsilon}},\hat{\underline{J}}_{BS}\}$ such that the
disc plane is described by the plane $\hat{\underline{J}}_{BS}=0$.
$\hat{\underline{J}}_{flow}$ is the
angular momentum vector of the flow and we see its orientation
precesses around the blue dotted ring, its phase described by the
precession angle $\gamma$. The flow, shown in (translucent) blue, is
then described by the plane orthogonal to
$\hat{\underline{J}}_{flow}$. The observer's position is described by
$\theta_i$ and $\phi_i$.}
\label{fig:schem}
\end{figure}

The geometry we consider for the two component accretion flow is
illustrated in Figure \ref{fig:schem}.  We assume that the disc has
angular momentum vector set by the binary system,
$\hat{\underline{J}}_{BS}$, and that this is misaligned with the spin
axis of the black hole (the z-axis) by an angle $\beta$.  The flow
angular momentum vector, $\hat{\underline{J}}_{flow}$, precesses
around the z-axis with phase given by the precession angle,
$\gamma$. The plane of the disc is the plane orthogonal to
$\hat{\underline{J}}_{BS}$, while the plane of the flow is 
orthogonal to $\hat{\underline{J}}_{flow}$. In this coordinate system,
the binary partner will orbit in the `disc' plane. The observer's
position is described by an inclination angle, $\theta_i$, and a
viewer azimuth, $\phi_i$, which can take the range of values
$0 \le \theta_i \le \pi/2$ and $0\le \phi_i \le 2\pi$. Here,
$\theta_i$ is defined with respect to the binary (i.e. the disc)
angular momentum vector and $\phi_i$ is defined with respect to
the x-axis.

The flow then precesses around a circle centred on the black hole spin
axis, from being aligned with the disc when $\gamma=0$, to being
misaligned by angle $2\beta$ with respect to the disc when $\gamma=\pi$.
We can define a vector $\hat{\underline{r}}_d$ which points from the
black hole to any point on the disc plane. If the top of the flow is
its brightest part, the region of the disc most strongly illuminated
by the flow for a given $\gamma$ is where the angle between
$\hat{\underline{r}}_d$ and $\hat{\underline{J}}_{flow}$ is smallest.
The smallest this angle can ever be is for $\hat{\underline{r}}_d
= \hat{\underline{\epsilon}}$ when $\gamma=\pi$; i.e. this is the
most that the flow angular momentum vector ever aligns with any
azimuth of the disc plane. $\hat{\underline{\epsilon}}$ therefore
defines the azimuth of the disc which sees the maximum illumination
from the flow. Material in the disc is spinning rapidly and, because
precession is prograde, this orbital motion is anti-clockwise for our
geometry. The viewer azimuth $\phi_i$ therefore specifies the
direction with respect to the viewer in which disc material in the
maximally illuminated region (i.e. on the $\hat{\underline{\epsilon}}$
axis) is moving. For $\phi_i=0$, the receding part of the disc is
most strongly illuminated as the flow precesses.  Instead, for $\phi_i
=\pi/2$ the maximum illumination is on the patch directly in front of
the black hole. For $\phi_i=\pi$ the maximum illumination is on the
approaching side of the disc, while for $3\pi/2$ it it for the patch
directly behind the black hole. We assume that the disc is razor thin
and flat (i.e. no flaring). The mathematical definitions for the
geometry we use are outlined in Appendix \ref{sec:geomcalc}.

\begin{figure}
\centering
\leavevmode  \epsfxsize=7.5cm \epsfbox{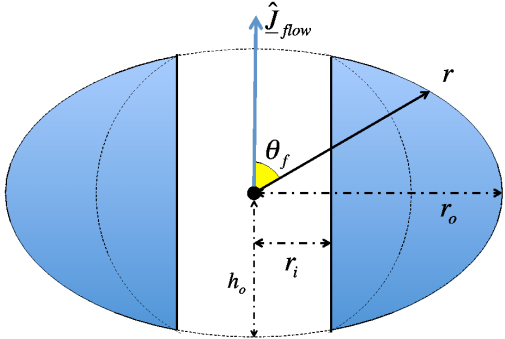}
\caption{Schematic diagram illustrating the cross section of the flow.
See text for details.}
\label{fig:ellipse}
\end{figure}

\subsection{Flow}
\label{sec:flow}

Unlike the disc, the flow has volume and scale height, so we must make
some assumptions about its shape.  We assume that it is a crushed
sphere; i.e. viewed from above it is circular but it has an elliptical
cross section as illustrated in Figure \ref{fig:ellipse}.  The
semi-major axis of the ellipse is $r_o$ and the semi-minor axis is
$h_o$.  We choose to parametrise this by defining a scaleheight,
$h/r$, such that $h_o=(h/r)r_o$. Figure \ref{fig:ellipse} also shows
that we set an inner radius, $r_i$, such that the core of the
quasi-spherical flow is missing. This is to incorporate a flavour of
the numerical simulations which show that shocks (at the bending wave
radius) can truncate the inner region of the hot flow (Fragile et al
2007).  Any point on the flow surface is then a distance $r$ away from
the black hole, where $r$ is a function of the angle $\theta_f$. We
assume that each radius of the surface radiates the gravitational
potential energy released at that radius (i.e. we use a surface rather
than a volume emissivity).  This gives a simple analytic model where
the central parts of the flow (outside of $r_i$) are brighter than the
outer parts, but that these bright regions are near the poles which
gives a reasonable reflection fraction, $(\Omega/2\pi)$ while also
giving a reasonable precession frequency (set by $r_i$, $r_o$, $M$,
the surface density profile which we assume to be constant, and $a_*$,
where $a_*$ is the dimensionless spin parameter: equation 1 in Ingram,
Done \& Fragile 2009). Note that, eventhough this is a simplified
prescription, the most influential aspect of the flow geometry is
where the brightest region lies.  In nearly all imaginable geometries,
this point lies at the pole of the flow (as it does for our geometry).
Thus our mathematically convenient assumptions for flow geometry should
provide us with results not materially different from a far more
difficult calculation assuming a geometry identical to the Fragile et al
(2007) simulation. More details of the flow geometry are presented
in Appendix \ref{sec:geomcalc}.

Fundamentally, the precession frequency modulates the continuum as the
pole moves in and out of sight. The
QPO maximum occurs when the pole faces the observer and the minimum
when it faces away. Thus the region of the disc preferentially
illuminated is in front of the black hole (from the point of view of
the observer) at the QPO maximum and behind for a QPO minimum.
Because precession is prograde, this means that the flow illuminates
the approaching disc material during the rise to a QPO maximum
(because the pole has to first move towards us in order to face us)
and the receding material on the fall to a QPO minimum. Below we
calculate the self-consistent illumination pattern for the disc as a
function of QPO phase for our assumed geometry.

\section{Implications of a precessing flow} \label{sec:model}

\subsection{Disc irradiation}
\label{sec:discirr}

\begin{figure}
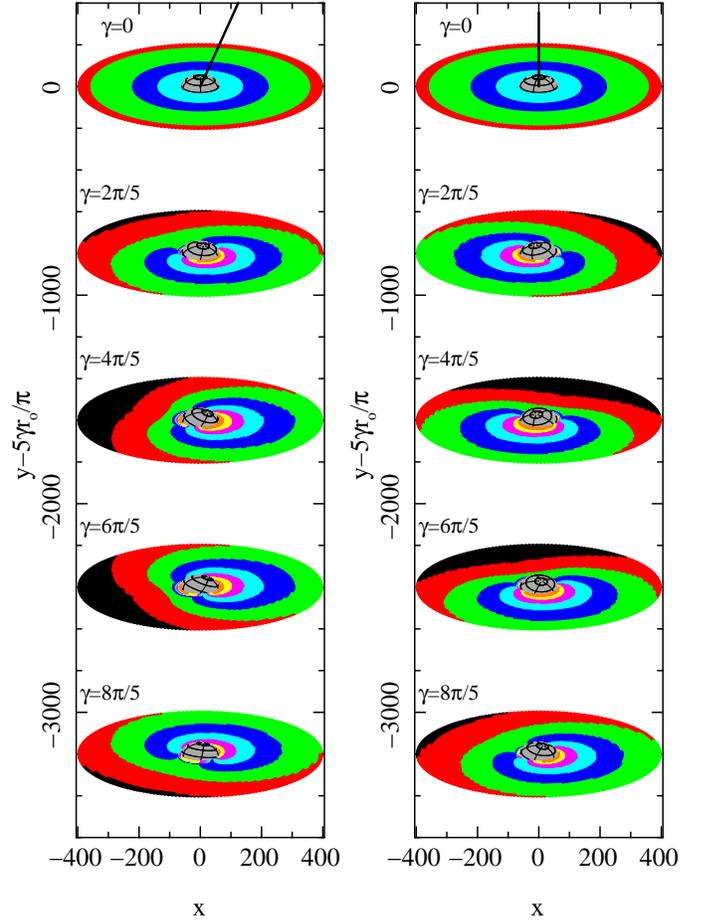

\centering$
\begin{array}{c}
\leavevmode  \epsfxsize=4.5cm \epsfbox{discpic_ro60_ri07_b15_hr090_phii000_thetai60.ps}
\leavevmode  \epsfxsize=4.5cm \epsfbox{discpic_ro60_ri07_b15_hr090_phii090_thetai60.ps}
\end{array}$
\caption{
Disc irradiation by the flow as seen by a viewer with $\theta_i=60^o$ and
$\phi_i=0^o$ (left) or $\phi_i=90^o$ (right). The flow is shown in grey with black
gridlines for clarity. The truncation radius is $r_o=60$. The luminosity incident on
the disc is grouped into 8 bins with black, red, green, blue, cyan, magenta, yellow
and orange representing the dimmest to brightest patches on the disc.
The solid black line in the top picture of each plot indicates the black hole spin
axis. Flow precession causes the characteristic illumination pattern to rotate around
the disc.
}
\label{fig:discpic}
\end{figure}

\begin{figure}
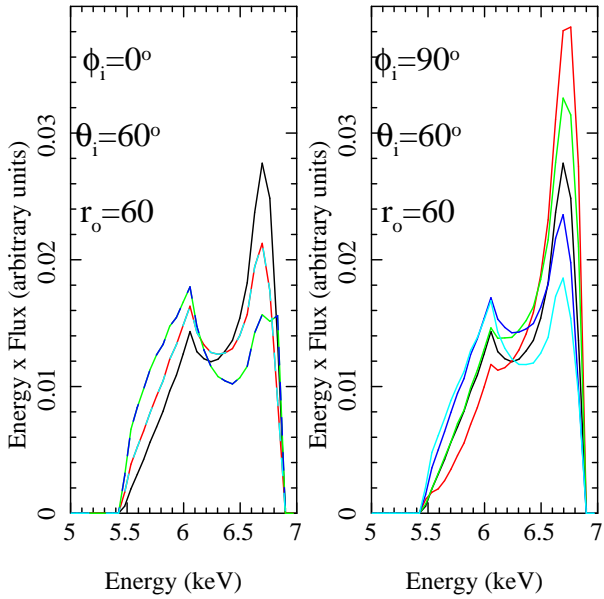

\centering$
\begin{array}{c}
\leavevmode  \epsfxsize=4.cm \epsfbox{delta_ro60_ri07_b15_hr090_phii000_thetai60.ps}
\leavevmode  \epsfxsize=4.cm \epsfbox{delta_ro60_ri07_b15_hr090_phii090_thetai60.ps}
\end{array}$
\caption{
The iron line profile as seen by a viewer with $\theta_i=60^o$ and $\phi_i=0^o$ (left)
or $\phi_i=90^o$ (right). The rest frame iron line profile is assumed to be a
$\delta-$function at $6.4$ keV and the truncation radius is $r_o=60$ as in
Figure \ref{fig:discpic}. Different colours represent different snapshots in time
with black, red, green, blue and cyan representing the top to bottom snapshots
pictured in Figure \ref{fig:discpic}. The rotation of the illumination pattern
causes the iron line profile to rock from red to blue shift.
}
\label{fig:delta}
\end{figure}

\begin{figure}
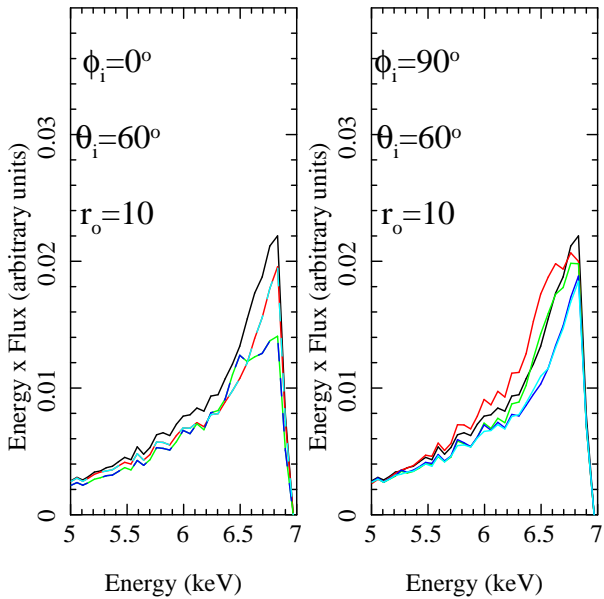

\centering$
\begin{array}{c}
\leavevmode  \epsfxsize=4.cm \epsfbox{delta_ro10_ri07_b15_hr090_phii000_thetai60.ps}
\leavevmode  \epsfxsize=4.cm \epsfbox{delta_ro10_ri07_b15_hr090_phii090_thetai60.ps}
\end{array}$
\caption{
The iron line profile as seen by a viewer with $\theta_i=60^o$ and $\phi_i=0^o$ (left)
or $\phi_i=90^o$ (right). The rest frame iron line profile is still assumed to be a
$\delta-$function at $6.4$ keV but the truncation radius is now $r_o=10$. The different
colours represent the same snapshots in time as in Figure
\ref{fig:delta}. We see the motion of the iron line is different here
compared to Figure \ref{fig:delta}. Due to stronger Doppler and relativistic boosting
in the inner disc, the red wing never dominates in the $E<6.4$ keV
region.
}
\label{fig:delta2}
\end{figure}

Each flow surface element will radiate a luminosity $dL$ over a
semi-sphere (because the element radiates \textit{away} from the black
hole). A disc surface element with area $dA_d$ will intercept some
fraction of this luminosity. This fraction can be calculated
self-consistently from the projected area of the disc element. The 
disc element will not intercept any of the
luminosity from the flow element if it makes an angle greater than
$\pi/2$ with a vector which is orthogonal to the flow element and
points away from the black hole (i.e. if it is not in the unit
semi-sphere of the flow element). Also, as observers with $\theta_i
\leq 90^o$, we only observe reflected photons which are intercepted by
the \textit{top} of the disc.

The total incident luminosity on the disc surface element is
calculated by integrating over the surface of the entire flow. We do
this calculation for every disc surface element over a full range of
precession angles ($0\leq \gamma <2\pi$) in order to build a picture
of disc irradiation as a function of precession angle (and therefore
time). The details of this calculation are presented in Appendix
\ref{sec:disccalc}. For simplicity, we use a Euclidean metric
i.e. assume that light travels in straight lines.  This should be a
fairly reasonable approximation because we assume a fairly
large value of $r_i$ throughout the paper (following Dexter \& Fragile
2011; Ingram \& Done 2012; Fragile 2009) and so lightbending is not
very significant (e.g. Fabian et al 1989).

Throughout the paper, we will use the values $r_i=7$, $\beta=15^o$ and
$h/r=0.9$ (we discuss our reasoning for these fiducial values in
section \ref{sec:continuum}). Figure \ref{fig:discpic} shows the
resulting illumination pattern with $r_o=60$, with snapshots taken at
five different values of precession angle $\gamma$ for an inclination
angle of $\theta_i=60^o$.  The left
hand plot shows the pattern as seen by an observer at $\phi_i=0^o$,
whereas the right hand plot shows this for $\phi_i=90^o$.  The
luminosity is grouped into bins of equal logarithmic size with black,
red, green, blue, cyan, magenta, yellow and orange representing the
dimmest to brightest bins respectively. The flow is shown in grey with
black gridlines included for clarity. In the top picture of each plot,
we also include a straight black line to illustrate the orientation of
the black hole spin axis. This is misaligned with
$\underline{\hat{J}}_{BS}$ by $\beta=15^o$ but, as Figure
\ref{fig:discpic} demonstrates, the apparent misalignment between
these two vectors depends on the viewing position. We clearly see the
flow precess, with the pole of the flow moving in a circle around the black
hole spin axis. As it does, the brightest part of the disc is always
the region closest to the pole of the flow meaning that it
\textit{rotates} around the disc.  Because of our asymmetric geometry,
the flow starts off aligned with the disc, is misaligned by $2\beta$
when $\gamma=\pi$ before aligning again for $\gamma=2\pi$. For
$\phi_i=0$, the maximum misalignment (giving the maximum illumination of
the disc) is on the right hand (receding) side of the disc, while for
$\phi_i=90$ it is directly in front of the black hole, but in both cases
the illumination pattern rotates. In the next section, we will discuss
how this  will affect the observed iron K$_\alpha$ line.

\subsection{Effect on the iron K$_\alpha$ line profile}
\label{sec:delta}

When the flow emission irradiates the disc, bound atoms in the disc
will fluoresce to produce emission lines, the most prominent being the
iron K$_\alpha$ line at $\sim 6.4$ keV (George \& Fabian 1991; Matt,
Perola \& Piro 1991). However, this line is in the rest frame of the
disc which is rotating rapidly meaning that a non face-on observer
will see some regions of the disc moving towards them and others
receding.  Doppler shifts mean that emission from the approaching side
is blue shifted while that on the receding side is red shifted. Also,
length contraction along the line of motion beams the emission in that
direction. Thus the blue shifted emission from the approaching side is
also boosted in comparison to the red shifted emission, leading to a
broadened and skewed iron line. An additional energy shift is provided
by time dilation and also gravitational redshift which combine to
broaden the line even further (Fabian et al 1989; 2000). Figure
\ref{fig:discpic} clearly shows that, according to this model, the
disc irradiation pattern \textit{rotates} around the disc meaning that
sometimes the brightest region of the disc is receding (e.g. the
$\phi_i=0^o$, $\gamma=4\pi/5$ scenario in Figure \ref{fig:discpic}),
and sometimes the brightest region is approaching (e.g. the
$\phi_i=90^o$, $\gamma=2\pi/5$ scenario in Figure
\ref{fig:discpic}). Therefore, as the flow precesses, the iron line
will periodically rock between red and blue shift.  In this example,
the material in the disc and the irradiation pattern are both rotating
anti-clockwise. In general, they they could both be moving clockwise
but the resulting pattern is the same (maximum blueshift, QPO maximum,
maximum redshift, QPO minimum). Lense-Thirring precession is prograde,
so the disc and flow will never be rotating in opposite directions,
making this periodic shifting of the iron line profile a unique
prediction of the model. 

We use the illumination pattern on each surface element of the disc to
set the amount of intrinsic iron line emission. We assume that this
is a $\delta-$function at $E_0 =6.4$keV and then use the radius and
azimuth of the surface element of the disc and the inclination of the
observer to calculate the shifted line emission (see Appendix).

Figure \ref{fig:delta} shows the iron line profile at five snapshots
of time with black, red, green, blue and cyan lines corresponding to
$\gamma=0$, $2\pi/5$, $4\pi/5$, $6\pi/5$ and $8\pi/5$ respectively.
We use the same parameters as for Figure \ref{fig:discpic}. The
details of this calculation are presented in Appendix
\ref{sec:ironcalc}. For simplicity, we do not include light bending
but this should not be a large effect for the comparatively
large radii we consider. The left hand plot is for $\phi_i=0^o$,
$r_o=60$ (i.e. corresponding to the left plot of Figure
\ref{fig:discpic}) and we see that the iron line does indeed rock
between red and blue shift as the illumination pattern rotates.  Note
that, for these parameters, the 2nd and 5th snapshots have an
identical iron line profile, as do the 3rd and 4th snapshots. The
right hand plot is for $\phi_i=90^o$, $r_o=60$ (i.e. corresponding to
the right hand plot in Figure \ref{fig:discpic}). We see that the
periodic rocking has a different phase and the peak flux of the blue
wing is much larger. This is because, for the $\phi_i=0^o$ case, the
approaching side of the disc is never the brightest part, whereas this
does happen for the $\phi_i=90^o$ case. This movement of the iron line
is obviously a very distinctive model prediction and so could provide
a detectable, unambiguous signature of a vertically tilted, prograde
precessing flow i.e. a clean test of a Lense-Thirring origin of the
QPO.

Figure \ref{fig:delta2} shows the same thing but now $r_o=10$. We see
that Doppler (and relativistic) boosting of the blue wing is now such
a large effect that
the red wing never dominates even when the flow is preferentially
illuminating the receding material. As such, the motion of the iron
line is different. Crucially, although the exact shape of the iron
line depends on the illumination pattern and thus the details of the
assumed flow geometry, this dependence on truncation radius is really
quite robust to changes in flow geometry. The differences between
Figures \ref{fig:delta} and \ref{fig:delta2} are driven primarily by
the difference in disc angular velocity (i.e. the position of the
truncation radius) and not the details of the modelling. Thus this
effect could provide a robust diagnostic for the accretion flow
geometry.

\subsection{Modulation of the continuum}
\label{sec:continuum}

As the flow precesses, the luminosity seen by the observer will change
periodically giving rise to a strong QPO (with the quasi-periodicity
provided by frequency jitter among other processes; Ingram \& Done
2012; Heil, Vaughan \& Uttley 2011; Lachowicz \& Done 2010).  This is
because the total surface area of the flow viewed by the observer
changes and, also, some regions of the flow are brighter than others
meaning that a trough in the light curve would typically occur when
the brightest regions of the flow (i.e. the poles) are hidden.  The
calculation for this process is similar to that performed in section
\ref{sec:discirr}.  Each flow surface element emits a luminosity
$dL$. The observer at $\theta_i$, $\phi_i$ will see no luminosity from
this surface element if they are not within the unit semi-sphere of
the element, and we also remove luminosity from lines of sight which
are obstructed by the disc. We can then integrate over every flow
element to calculate the observed luminosity as a function of
precession angle.

\begin{figure}
\centering
\leavevmode  \epsfxsize=7.5cm \epsfbox{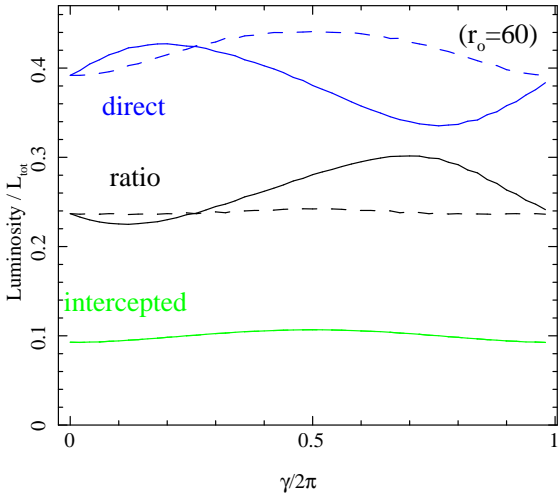}
\caption{
Emission as seen by a viewer at $\theta_i=60^o$ and $\phi_i=0^o$ (solid lines)
or $\phi_i=90^o$ (dashed lines). The blue line represents emission directly
observed from the flow. We see that precession of the flow introduces a strong
periodicity. The green line represents the total luminosity intercepted by the
disc. This also has a periodicity because the misalignment between disc and flow
changes as the precession angle, $\gamma$, evolves. It does not, however, depend
on the position of the observer. The black line is the ratio between direct and
reflected (intercepted) light, ($\Omega/2\pi$).
}
\label{fig:ratio}
\end{figure}

The blue lines in Figure \ref{fig:ratio} show the observed luminosity
expressed as a fraction of the total luminosity, $L_{tot}$, plotted
against precession angle. We use the fiducial parameters
$r_i=7$, $\beta=15^o$ and $h/r=0.9$ and consider the $r_o=60$ example.
The solid line is for $\phi_i=0^o$ and the dashed line represents
$\phi_i=90^o$. As expected, the observed luminosity varies with
precession angle and the phase depends on $\phi_i$. The fractional
rms is $8.4\%$ and $4.2\%$ for $\phi_i=0^o$ and $\phi_i=90^o$
respectively. These values are lower than the observed QPO rms values
of $\sim 10-15\%$. However, the predicted values would be higher if we
were to consider that the flow is fed by disc photons, the flux of
which incident on the flow will change periodically as the flow
precesses. We ignore this process here because it will affect the
direct and reflected emission equally and so will not contribute to
the rocking iron line effect.

For the green line, we plot the total luminosity incident on the disc
(which determines the iron line / reflected flux) as a function of
precession angle. Because the disc is flat, this does not depend on
$\phi_i$.  This effectively tracks the misalignment between flow and
disc with the minimum reflection occurring when the flow is aligned
($\gamma=0$) and the maximum when the flow is misaligned by $2\beta$
($\gamma=\pi$). Hence the direct and intercepted
emission are generally out of phase. The black lines show the reflection
fraction (intercepted/direct) with the solid and dashed lines
representing $\phi_i=0$ and $90^o$ respectively. 
This corresponds to the solid angle of the disc, and the time averaged
ratio for $\phi_i=0^o$ is $\Omega/2\pi=0.263$, and 
with $\Omega/2\pi=0.238$ for the $\phi_i=90^o$ case. These values are
fairly representative of those observed for the low/hard state
(e.g. Gierlinski et al 1999; Zycki, Done \& Smith 1998; Gilfanov 2010).

Note that
large value of $h/r$ gives a reasonable reflection fraction but
under predicts the QPO rms. If we had considered, for example, an
overlap region between disc and flow, disc flares or a small disc scale
height, we could have achieved a
reasonable reflection fraction \textit{and} the correct QPO rms
(for this we would also need to consider the variation in disc seed
photons) for a far lower value of $h/r$. However, these effects are
all very difficult to model and our assumed geometry should not
significantly affect the final results. Thus we choose the fiducial
parameter values to give reasonable results for a simplified
geometry.

\section{Spectral modelling}
\label{sec:spec}

We now use a full reflected spectrum rather than just a line, and
recalculate the effect of the  rotating disc
illumination pattern and varying effective area of the flow for this
more realistic scenario. We consider the same two values of truncation
radius as those considered previously, $r_o=60$ and $r_o=10$. These
values correspond to precession frequencies of $f_{QPO}(r_o=60) =
0.145$ Hz and $f_{QPO}(r_o=10) = 5.36$ Hz for the fiducial parameters,
a spin of $a_*=0.5$ and a mass of $M=10M_{\sun}$ (i.e. $2.9$ and
$107.1$ $a_*(M_{\sun}/M)$ Hz). They also correspond to different
spectral states, with $r_o=60$ giving rise to a low/hard state (LHS)
spectrum and $r_o=10$ leading to a soft intermediate state (SIMS)
spectrum. The QPO in the LHS spectrum will be of type-C whereas it
will be of type-B for the SIMS spectrum.

\subsection{Method}
\label{sec:method}

For both the LHS and SIMS spectra, we include quasi-thermal disc
emission, Comptonised flow emission and a reflection spectrum. We use
\textsc{xspec}v12 (Arnaud 1996) throughout. 
We describe the disc with {\sc diskbb} (Mitsuda et al 1984), 
and for simplicity we assume
that this spectrum is constant. This is not strictly true. 
Figure \ref{fig:discpic} shows that the inner disc is
periodically obstructed by the flow, giving a small periodicity in the
hottest part of the disc emission. Also, the non-reflected photons
which illuminate the disc will thermalise and add to the intrinsic
disc emission, and this additional thermal emission will vary in
intensity, being stronger when the flow is at its maximum misalignment
angle to the disc, and weakest when the flow is aligned with the
disc. This additional thermal emission is also periodically
redshifted/blueshifted in the same way as the line. However, these
effects should be small as they are diluted by the much larger
constant flux from the disc. We will investigate this in a future
paper, as evidence for this may have been observed (Wilkinson
2011). However, here we are interested in the iron line region and so
ignore this potential contribution to the QPO in the disc spectrum.

For the flow we assume
that every element emits the same spectrum, meaning that the
periodicity is in the normalisation of the flow spectrum. 
We describe the
spectrum by the Comptonisation model {\sc nthcomp}
(Zdziarski, Johnson \& Magdziarz
1996; Zycki, Done \& Smith 1999) which produces a power law spectrum
with high and low energy cut-offs governed by the electron
temperature and disc photon temperature ($kT_{bb}$ tied to 
the disc temperature) respectively. We fix the normalisation of this
by the angle averaged flux from the flow ($L_{tot}$), to set the flux
from each surface element of the flow. We then use the
method described in section \ref{sec:continuum} to determine the
modulation of the observed continuum, to calculate the factor by
which to multiply the normalisation of {\sc nthcomp} as a function
of phase angle. 

We use the method described in section \ref{sec:delta} to calculate
the illuminating flux from the flow at each surface element in the
disc, and use this to set the normalisation of the illuminating {\sc
nthcomp} model.  We describe the shape of the resulting reflected
emission by {\sc rfxconv} (Ross \& Fabian 2005; Done \& Gierlinski
2006; Magdziarz \& Zdziarski 1995; Kolehmainen, Done \& Diaz Trigo
2011). This is similar in form to the {\sc ireflect} model in {\sc xspec}
but replaces the very approximate ionisation balance incorporated in
this model with the much better Ross \& Fabian (2005) calculations.
This outputs a partially ionised (parametrised by $\log_{10}{\xi}$)
reflection spectrum, including the self consistent emission lines, for
a general illuminating spectrum. We fix the inclination angle of the
reflector at $\theta_i$ and abundances at solar.  We calculate the
reflected emission from this illuminating flux assuming
$\Omega/2\pi=1$. This is an underestimate as {\sc rfxconv} assumes
that the disc is illuminated isotropically, whereas in our geometry
the illumination is preferentially at grazing incidence.  However, the
amount of reflection is also set by the unknown details of the shape
of the flow, so this approximation is good enough to demonstrate the
general behaviour of the model.

The reflected emission from each surface element is 
shifted  in energy depending on the radius and azimuth (see Appendix
\ref{sec:ironcalc}).
We sum the reflected emission from all the disc elements 
to derive the total reflected emission for each phase. 
This gives the correct relative 
normalisation of the continuum and reflected flux, and 
how this changes as a function of precession phase angle $\gamma$
for a given set of model 
($r_o$, $r_i$, $\beta$, $h/r$, $\theta_i$, $\phi_i$) and spectral
($kT_{bb}$, $\Gamma$, $\log\xi$, $kT_e$) parameters.

\subsection{Phase resolved spectra}
\label{sec:phaseres}

\begin{table}
\centering
\begin{tabular}{l|l|l|l|c}
 \hline
 $ $                  & $ $                 & LHS              & SIMS               \\
 \hline
 \hline
\textsc{phabs}        & $N_h$ (cm$^{-2}$)   & $1\times 10^{22}$ & $1\times 10^{22}$\\
\hline
\textsc{diskbb}       & $kT_{bb}$ (keV)     & $0.1$             & $0.5$            \\
 $ $                  & norm                & $1\times 10^8$    & $5\times 10^4$   \\
\hline
\textsc{nthcomp}      & $kT_{bb}$ (keV)     & $0.1$             & $0.5$            \\
 $ $                  & $kT_{e}$ (keV)      & $100$             & $60$             \\
 $ $                  & $\Gamma$            & $1.7$             & $2.4$            \\
 $ $                  & norm                & $5$               & $4$              \\
\hline
\textsc{rfxconv}      & $\Omega/2\pi$       & $1$               & $1$              \\
 $ $                  & $\log_{10}{\xi}$    & $2.4$             & $3$              \\
 $ $                  & norm                & $5$               & $4$              \\
\hline
QPO                   & $r_o$ ($R_g$)       & $60$              & $10$             \\
modulation            & $\beta$ (degrees)   & $15$              & $15$             \\
\&                    & $r_i$ ($R_g$)       & $7$               & $7$              \\
smearing              & $h/r$               & $0.9$             & $0.9$            \\
\hline
\end{tabular}
\caption{Summary of the parameters used for both the LHS and SIMS spectral models.}

\label{tab:parameters}
\end{table}

\begin{figure}
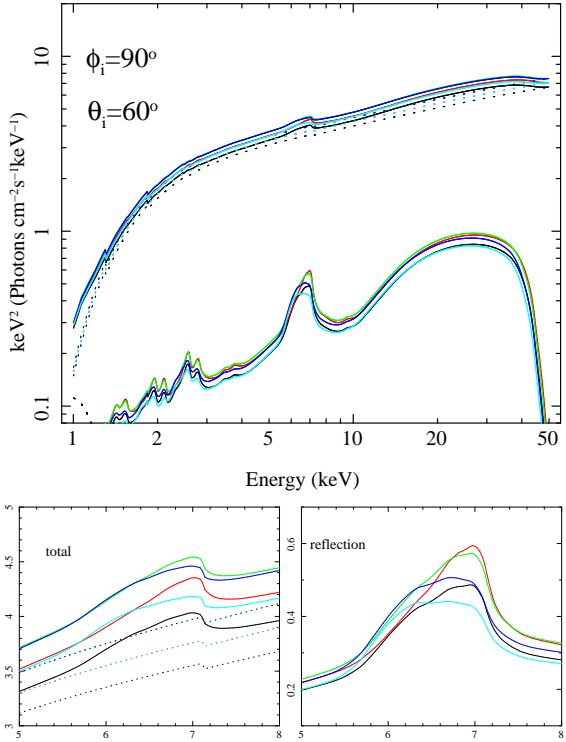

\centering$
\begin{array}{c}
\leavevmode  \epsfxsize=7.5cm \epsfbox{tot_lhs.ps} \\
\leavevmode  \epsfxsize=3.75cm \epsfbox{closeup_tot_lhs.ps}
\leavevmode  \epsfxsize=3.75cm \epsfbox{closeup_ref_lhs.ps}
\end{array}$
\caption{
LHS spectrum for five snapshots in time calculated using the model described in
the text, using the parameters listed in Table \ref{tab:parameters}.
We use the same convention as for Figures \ref{fig:delta} and \ref{fig:delta2} with
black, red, green, blue and cyan representing the first to last snapshots. The top
plot is a broadband spectrum with all of the components. The disc and Comptonisation
components are both represented by dotted lines and the total spectrum as well as the
reflection component are represented by solid lines. The bottom right plot zooms in
on the intrinsic iron line and the bottom left plot zooms in on the iron line region
of the total spectrum. We see that the motion of the iron line is still present but
dilution from the continuum makes the effect much more subtle in the total spectrum.
}
\label{fig:speclhs}
\end{figure}

\begin{figure}
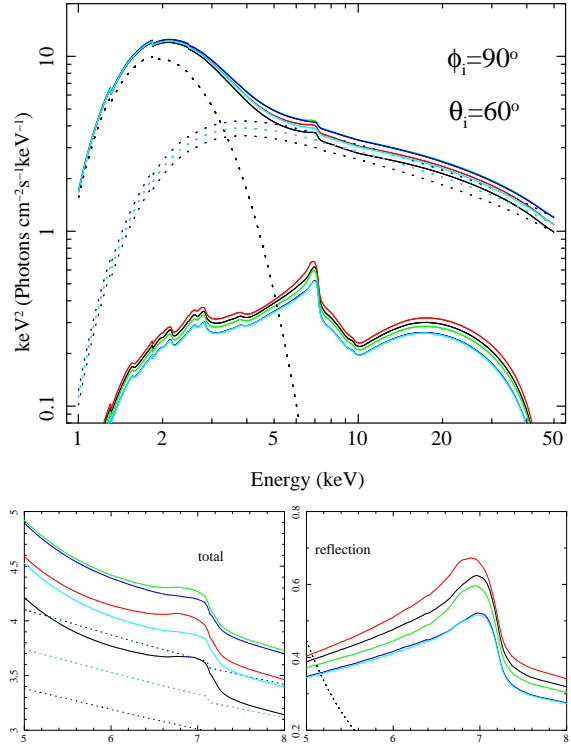

\centering$
\begin{array}{c}
\leavevmode  \epsfxsize=7.5cm \epsfbox{tot_sis.ps} \\
\leavevmode  \epsfxsize=3.75cm \epsfbox{closeup_tot_sis.ps}
\leavevmode  \epsfxsize=3.75cm \epsfbox{closeup_ref_sis.ps}
\end{array}$
\caption{
SIMS spectrum for five snapshots in time calculated using the model described
in the text, using the parameters listed in Table \ref{tab:parameters}.
We use the same conventions as for Figure \ref{fig:speclhs}. We see that, as
for the $\delta-$function calculation, the movement of the iron line
is characteristically different for the SIMS compared with the LHS.
}
\label{fig:specsis}
\end{figure}

The parameters used for each state are shown in Table \ref{tab:parameters}. We 
assume that $kT_{bb}$, $\Gamma$ and disc ionisation increase as the rise to
outburst continues whereas $kT_e$ decreases, as is commonly observed. The
resultant time averaged LHS ($r_o=60$) spectrum  has a $2-10$ keV flux of
$\sim 0.3$  Crab and $\Omega/2\pi=0.24$ (or iron line equivalent with of 150~eV
when fit by a diskline profile rather than a full reflected spectrum). 
For the SIMS  ($r_o=10$) spectrum, the flux is $\sim 0.66$ Crab and  the
reflection has $\Omega/2\pi=0.42$ (iron line equivalent width of $\sim 240$ eV)
with a much steeper continuum.  These values are typical of
those observed in the relevant states for fairly bright BHBs
(e.g. GRS1915+104 in its QPO state: Ueda et al 2010, and the
intermediate state of GX339-4; Tamura et al 2012), justifying our
choice of parameters.

Figure \ref{fig:speclhs} shows the LHS spectrum as viewed from a position with
$\phi_i=90^o$ and $\theta_i=60^o$ at five different snapshots in time. We use the
same convention as for Figures \ref{fig:delta} and \ref{fig:delta2} with black, red,
green, blue and cyan representing $\gamma=0$, $2\pi/5$, $4\pi/5$, $6\pi/5$ and
$8\pi/5$ respectively. The top plot shows the total spectrum (upper solid lines)
and its components, the constant disc (black dotted line just seen in the lower
left hand corner of the plot), variable flow (dotted continuum lines just underneath the total
spectra - the symmetry means that the red dotted line is the 
same as the cyan, while the
green is the same as the blue) and reflected spectra (lower solid lines).
We clearly
see the flow continuum oscillate while the reflection spectrum rocks between red and
blue shift, as well as changing in normalisation. The reflection spectrum is in
phase with the continuum in this example because $\phi_i=90^o$ (see Figure
\ref{fig:ratio}) but, in general, there is a phase difference between the two
components. The lower left plot zooms in on the iron line region in the total
spectrum, while the lower right plot shows the changes in the reflected emission.
We see that the reflected spectrum displays similar behaviour to the
corresponding $\delta-$function (right hand plot of Figure \ref{fig:delta}). The
rocking movement in the underlying reflection spectrum is still visible in the total 
spectrum, though somewhat diluted by the changing continuum level. 

Figure \ref{fig:specsis} shows the same thing but for the SIMS. As for the
$\delta-$function iron line profile in section \ref{sec:delta}, we see that
the major effect is now the strength and position of the blue wing rather than
a rocking motion from blue to red due to the much stronger Doppler and relativistic
boosting in the inner disc. Non theless, there is still a clear periodic shift in the line
shape with QPO phase, although the pronounced rocking of the iron line peak
energy predicted for the LHS provides more of a `smoking gun' for the Lense-Thirring
model.

\section{Observational predictions}
\label{sec:obs}

In this section, we consider how this effect may be best observed. One potential
method is to look at phase lags between different energy bands. We
could define a red wing energy band (say $5.4-6.4$keV) and a blue wing
energy band (say $6.4-7.4$keV) and look for a phase lag between the
two. However, Figures \ref{fig:speclhs} and \ref{fig:specsis} show
that, due to dilution from the periodically varying continuum, the
energy shifting of the iron line is very subtle in the total spectrum. This means
that the phase lag between red and blue wings is very small
($2-6\times 10^{-2}\pi$) for our model and, consequently may be
difficult to observe. Instead, we consider phase resolved spectroscopy.

\begin{figure}
\centering
\leavevmode  \epsfxsize=7.5cm \epsfbox{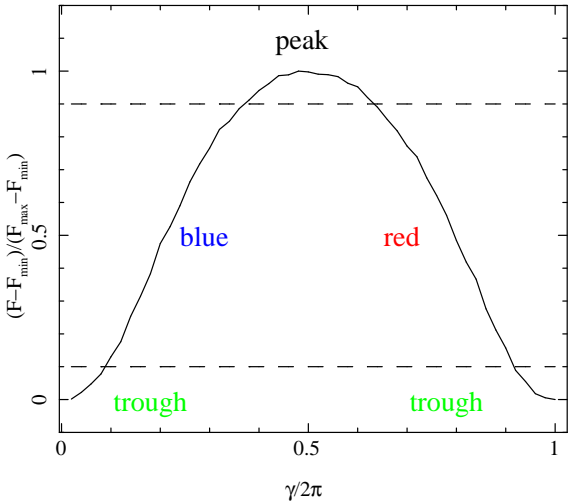}
\caption{The 2-20 keV integrated flux of the LHS model with
  $\phi_i=90^o$ and $\theta_i=60^o$ plotted against precession
  angle. The dashed lines are flux thresholds. Intervals of the light
  curve above the top dashed line are considered to be the QPO peak,
  intervals below the bottom dashed line are considered to be the
  trough. The rising section which will always follow a trough will
  have the bluest iron line profile. The falling section which always
  follows the peak will have the reddest iron line profile.}
\label{fig:sel}
\end{figure}

\begin{figure}
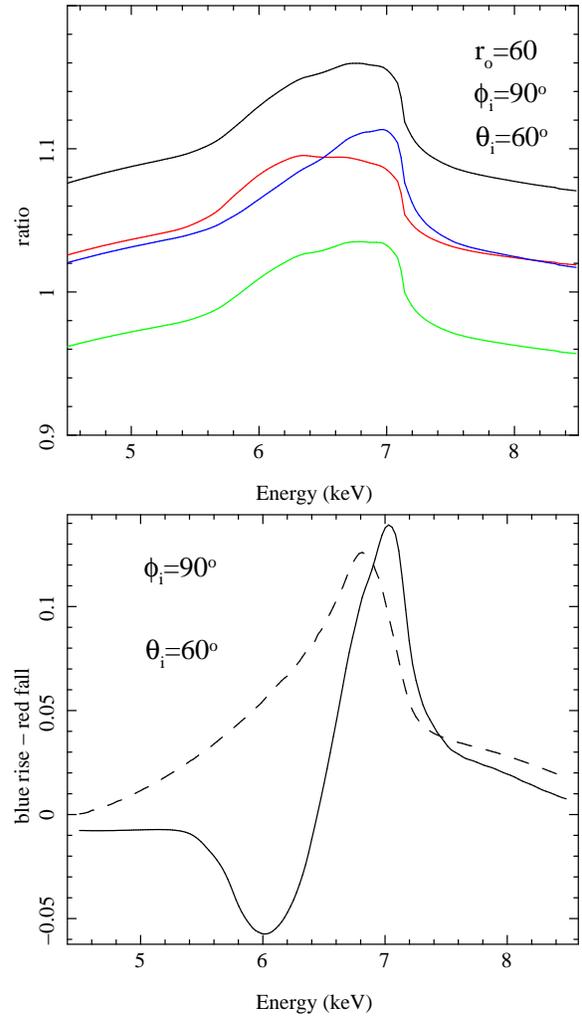

\centering$
\begin{array}{c}
\leavevmode  \epsfxsize=7.5cm \epsfbox{mod_sub.ps} \\
\leavevmode  \epsfxsize=7.5cm \epsfbox{diff_compro.ps}
\end{array}$
\caption{\textit{Top:} Phase binned spectra calculated assuming
$r_o=60$, $\phi_i=90^o$ and $\theta_i=60^o$ plotted as a ratio to
a power law with photon index $\Gamma=1.6$. These four phase bins
are for the QPO minimum (green), rise (blue), maximum (black) and fall
(red). As expected, the rise has the most heavily blue
shifted iron line and the fall has the most heavily red shifted iron
line.
\textit{Bottom:} The red fall spectrum subtracted from the blue rise
spectrum. The solid line is for the $r_o=60$ example shown in the top
plot and the dashed line is for $r_o=10$. The shape of this difference
spectrum is different for the two truncation radii. There is no
negative section in the dashed line because strong Doppler and relativistic
boosting in the inner disc prevents the red wing from dominating.}
\label{fig:sub}
\end{figure}

\subsection{Phase binning}
\label{sec:diff}

The random phase jumps and varying period characteristic of QPO light curves
make phase resolved spectroscopy difficult. Naively folding the light curve
on the QPO period is not appropriate. It is, however, possible to isolate
the maximum and minimum phase bins of the QPO by averaging over the brightest
and faintest points in the light curve. Miller \& Homan (2005) did this for
two GRS 1915+105 light curves, both containing a strong type-C QPO. This
allowed them to compare the spectra corresponding to the QPO peak and trough.
This analysis can be taken a step further because a rise will always follow
a trough and a fall will always follow a peak. This simple phase binning can
therefore provide four phase bins as opposed to two. Crucially, our model
predicts that the maximum red shift \textit{always} follows the QPO peak and
the maximum blue shift \textit{always} follows the QPO trough. This is because
the pole of the flow (which is the brightest region) faces us, then is moving
away from us, then faces away from us, then is moving towards us (before facing
us again). Therefore the flow illuminates the observer, then the receding (red
shift) part of the disc, then the region hidden to the observer then the
approaching (blue shift) part of the disc.

Figure \ref{fig:sel} shows the 2-20keV light curve of our LHS model with
$\phi_i=90^o$ and $\theta_i=60^o$.  We define a peak as the brightest
$10\%$ of the light curve and a trough as the faintest $10\%$. These
thresholds are shown as dashed lines. We can therefore isolate
the trough, the blue rise, the peak and the red fall. This flux selection
means that the majority of the counts lie in the more interesting rise
and fall sections as opposed to the peak and trough (unlike the flux
selection of Miller \& Homan who were interested in the peak and
trough spectra). Figure \ref{fig:sub} (top) shows the result of
averaging spectra belonging to each of these four phase bins. The
green line is the trough spectrum, the blue line is the rise spectrum,
the black line is the peak spectrum and the red line is the fall
spectrum. All are plotted as a ratio to a power law with photon index
$\Gamma=1.6$. We use this photon index rather than $\Gamma=1.7$
because the reflection hump makes the total spectrum harder than the
underlying Comptonisation. As expected, the rise spectrum contains the
most heavily blue shifted iron line and the fall spectrum contains the
most heavily red shifted iron line. Because we tie the normalisation
of the power law across the four spectra, we can see that the
peak spectrum has the highest flux, the trough spectrum has the lowest
and the rise and fall have comparable flux. 

In the bottom plot of Figure \ref{fig:sub}, we plot the red fall spectrum
subtracted from the blue rise spectrum. We use the absolute spectrum
in units of energy $\times$ flux rather than a ratio to a power
law. The solid line is for the example shown in the top plot where
$r_o=60$ and the dotted line is for $r_o=10$. When $r_o=60$, the red
wing of the iron line dominates during the fall meaning that the solid
line in the bottom plot dips below zero for $5.4 \gtrsim E \gtrsim
6.4$. During the fall, the blue wing dominates which gives rise to the
hump in the $6.4 \gtrsim E \gtrsim 7.4$ region. Due to Doppler (and relativistic)
boosting, the blue hump is larger than the red dip. When $r_o=10$, the
inner regions of the disc are moving much faster than the $r_o=60$
case and therefore boosting is a much more significant
effect. So much so, in fact, that the red wing of the iron line never
dominates over the blue wing, even during the fall. The dotted line in
the bottom plot therefore contains no red dip but only a blue hump.
The peak of the blue hump is lower for $r_o=10$ than for $r_o=60$ but
the area under the line is greater. This is because the iron line is
more heavily smeared in the $r_o=10$ case, again due to faster orbital
motion closer to the black hole. 

For both the LHS and the SIMS, the difference in iron line profile between the
QPO rise and the QPO fall is significant, offering the possibility of direct
observation for a range of spectral states. Note that this association of the
rise with the bluest profile and the fall with the redest profile is robust
as long as we are confident that the top (pole) of the flow is brighter than
the sides. Because type-B QPOs provide a far
cleaner signal than type-C QPOs, which are always coincident with broad band
variability, it will be easier to observe this effect for a source in the SIMS.
However, the QPO phase dependence of the iron line is particularly distinctive
for the LHS model. An enhanced blue wing on the QPO rise, as predicted for the
SIMS model, may feasibly be produced by some other process. A dominant red wing
on the QPO fall and an enhanced blue wing on the rise, as predicted for the LHS
model, can only realistically be produced by precession and a large
truncation radius. Moreover, an observation showing that the difference spectrum
\textit{changes} between states as we predict (i.e. the bottom plot of Figure
\ref{fig:sub}) would surely provide excellent evidence, not only of the
precession model, but also that the truncation radius moves between the LHS and
the SIMS. In the next section, we assess the likelihood of achieving such
observational confirmation.

\subsection{Simulated observations}
\label{sec:sim}

\begin{figure*}
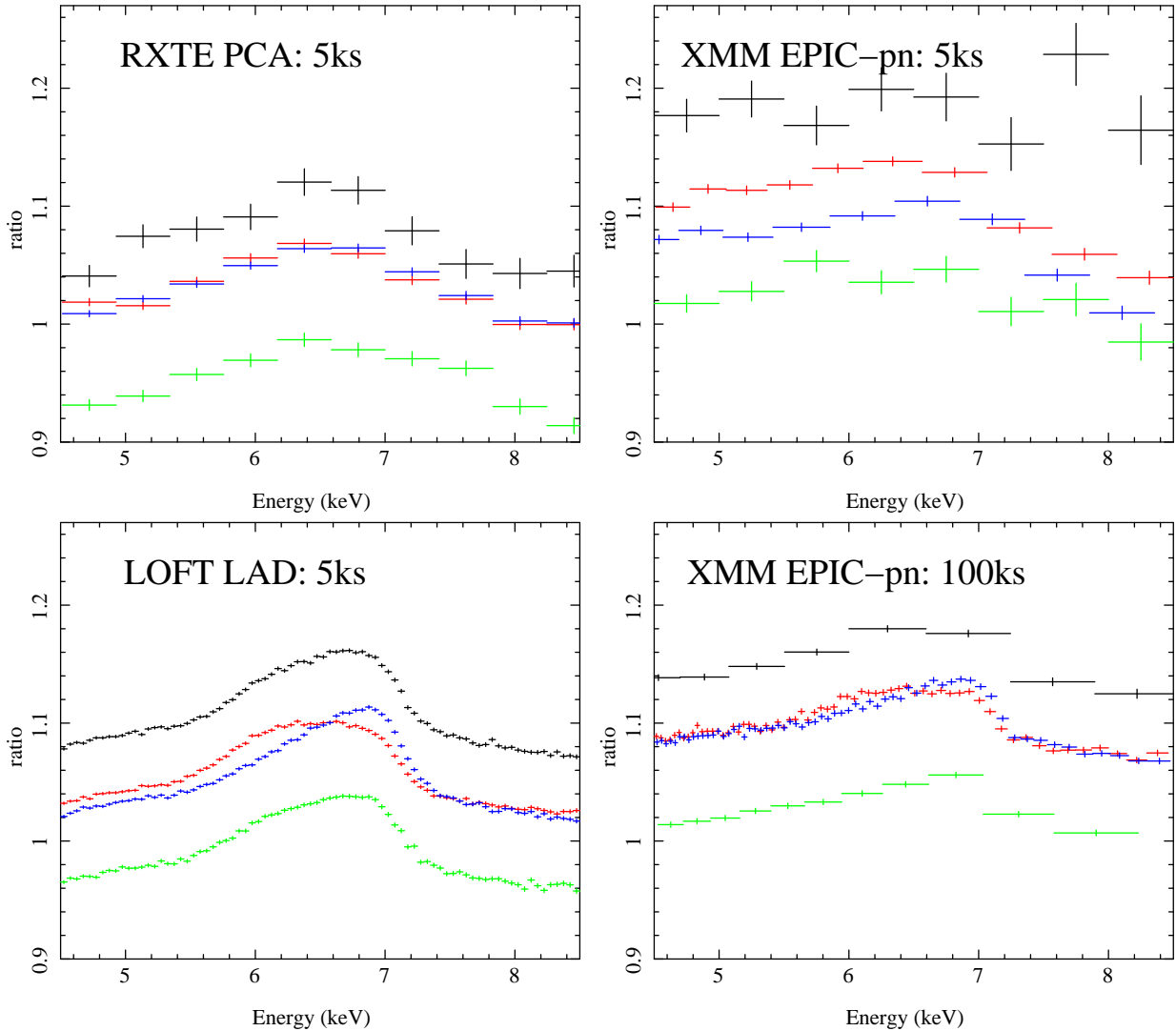

\centering$
\begin{array}{c}
\leavevmode  \epsfxsize=8.0cm \epsfbox{pca_sub.ps} ~~~~
\leavevmode  \epsfxsize=8.0cm \epsfbox{xmm_sub.ps} \\
\leavevmode  \epsfxsize=8.0cm \epsfbox{lad_sub.ps} ~~~~
\leavevmode  \epsfxsize=8.0cm \epsfbox{100ks_xmm.ps}

\end{array}$
\caption{Simulated observations of the phase binned spectra shown in
  Figure \ref{fig:sub} with $r_o=60$, $\phi_i=90^o$ and
  $\theta_i=60^o$. These spectra are unfolded around a flat power law
  and plotted as the ratio to a power law with $\Gamma=1.6$ and unity
  normalisation. Again the four phase bins are for the QPO minimum
  (green), rise (blue), maximum (black) and fall (red). Observed with
  the \textit{RXTE} PCA or the \textit{XMM Newton} EPIC-pn, it is
  difficult to see by eye the difference in iron line peak energy between
  different phase bins. In contrast, the \textit{Athena} WFI recovers the
  model well and the \textit{LOFT} LAD does so with an exceptionally high
  precision.}
\label{fig:ladandpca}
\end{figure*}

We test the feasibility of observation directly by simulating phase resolved spectra
using the \textsc{ftool} \textsc{fakeit}. This adds Poisson noise to a model
before subtracting a representative background and deconvolving around
a given response matrix. We simulate LHS spectra for 50 phase bins evenly spaced
in precession phase angle, $\gamma$. We assume
100s exposure for each phase bin. 
This corresponds to 50 $\times$ 100s = 5ks of good time. We sort the simulated data
into four phase bins just as we did with the model. For the simulated data, there is
just one QPO cycle with a long exposure but, for observational data there will be many
short exposure QPO cycles to average over. As long as any fluctuations in the
accretion geometry over this time are varying around an average value, the two
processes should be equivalent to a good approximation. 

The top left plot in Figure \ref{fig:ladandpca} shows the result of simulating
the response of the Rossi x-ray timing explorer (\textit{RXTE}) proportional
counter array (PCA; top layer, detector 2). We unfold the spectrum around a flat power
law and, as for the model, take the ratio to a power law with photon index
$\Gamma=1.6$. We use the same model as that shown in the top plot of
Figure \ref{fig:sub}; i.e. $r_o=60$, $\phi_i=90^o$,
$\theta_i=60^o$. Again, the green points are the trough, the blue
points are the rise, the black points are the peak and the red points
are the fall. Although a shift in line energy is visible between the rise
and fall spectra, it is unlikely to be statistically significant due to a
high noise level and low spectral resolution. The two
observations of GRS 1915+105 studied by Miller \& Homan (2005) were
both observed with \textit{RXTE} and, as such, the data were of a
comparable quality to our simulation. They fit the QPO peak and trough spectra
with a simple continuum model plus a Gaussian function for the iron line. When allowed
to be free in the fits, the centroid energy of the Gaussian was higher for the trough
spectrum than for the peak spectrum in both observations. However, they were also able
to achieve statistically acceptable results by fixing the centroid energy to the value
measured for the total spectrum. Therefore, although there is some evidence that the
line energy shifts, it is by no means statistically significant. It should be possible
to achieve a slightly more significant result with \textit{RXTE} data by comparing
the rise and fall phases rather than the peak and trough, but this is always
marginal in practice due to the limited energy resolution of
\textit{RXTE} fast timing modes.

The top right plot of Figure \ref{fig:ladandpca} shows the same thing
but for the \textit{XMM Newton} European photon imaging pn camera (EPIC-pn).
The Poisson noise level seems to be marginally worse compared with the
simulated PCA data. Although the spectral resolution of the EPIC-pn
is far better than that of the PCA, its effective area is less ($\sim$
0.05m$^2$ compared with $\sim$ 0.12m$^2$) meaning that we require a
very heavy re-binning to get a reasonable signal to noise. Therefore,
it may prove difficult to observe this effect using either \textit{RXTE}
or \textit{XMM Newton}. However, the number of counts in the rise and
fall phase bins could be maximised by halving the peak and trough
phase bins and adding them to either the rise or the fall (i.e. the
first half of the peak phase becomes part of the
rise and the second half becomes part of the fall). 

A longer exposure is required to reduce the counting errors. In the
bottom right hand panel of Figure \ref{fig:ladandpca}, we plot the
result of assuming a 100ks exposure for the EPIC-pn. Encouragingly, we
see that the dominant red wing in the falling phase is indeed
resolved. However, over such a long exposure time, parameters such as
$r_o$ may have systematically moved and so care must be taken to take
this into consideration.

The size of the effect is also
dependent on our assumptions. A smaller flow scaleheight would
increase the size of this effect because the flux emitted from the
poles of the flow would be an even greater fraction of the flux
emitted from the entire flow. Frame dragging could therefore have a
larger effect on the iron line than we predict here making it easier
to observe with current instruments than our simulations
imply. However, it also must be noted that the continuum will be more
complicated than we assume here with some QPO phase dependent spectral
pivoting resulting from a variation in the flux of disc photons incident on
the flow. This will make observation harder.

The bottom left hand plot of Figure \ref{fig:ladandpca} shows the
potential impact of the proposed mission \textit{LOFT} (the large
observatory for x-ray timing). We use the `required' response of
the large area detector (LAD), which is the principle instrument
of the mission. Because the LAD has an exceptionally large effective
area (10-12m$^2$), the results are far clearer than those provided
by current missions. In fact, the noise level is so
low with \textit{LOFT}, it would be possible to constrain spectra for
far more than four phase bins. We could also constrain these spectra
for less than 5ks good time, meaning that we could conduct detailed
studies of the evolution of the phase resolved spectra.

\subsection{RMS spectum}
\label{sec:rms}

\begin{figure}
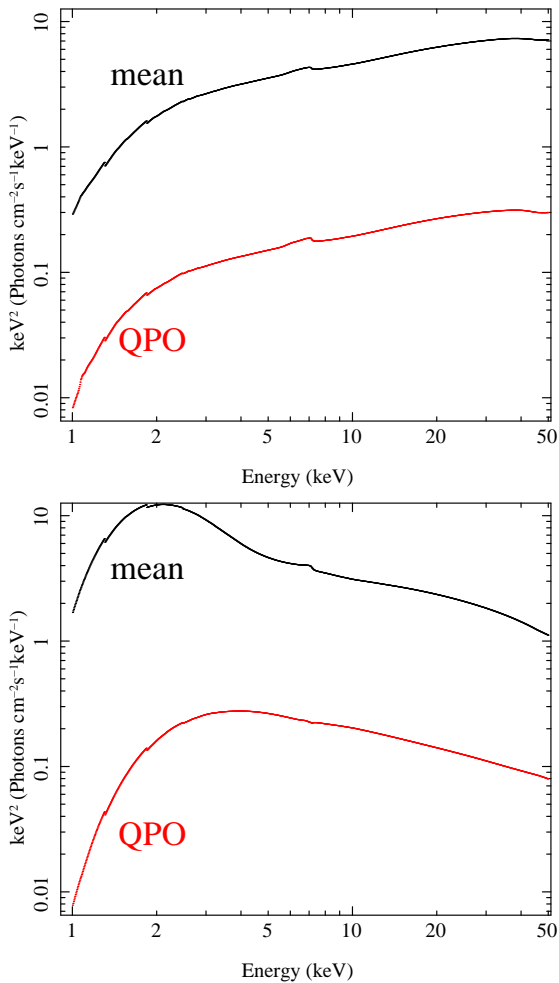

\centering$
\begin{array}{c}
\leavevmode  \epsfxsize=7.5cm \epsfbox{rms_lhs_fid.ps} \\
\leavevmode  \epsfxsize=7.5cm \epsfbox{rms_sis_fid.ps}
\end{array}$
\caption{Mean and QPO spectra for the LHS (top) and SIMS (bottom)
models. The QPO spectum is calculated by measuring the standard
deviation of each energy channel around the mean value across 50
values of precession angle.}
\label{fig:rms}
\end{figure}

Since we calculate 50 spectra for both the LHS and SIMS models, it
is simple to calculate the rms spectrum of the QPO. This is simply
the standard deviation of each energy channel in absolute units (i.e.
not divided through by the average). Figure \ref{fig:rms} shows this
for the LHS model (top) and the SIMS model (bottom) with the mean
spectrum plotted in black and the QPO spectrum plotted in red. Since
the QPO spectrum is fairly sensitive to model assumptions, it provides
a good way to constrain model parameters against observation. For the
models we use here, the misalignment angle $\beta$ is large and thus
we see reflection features in the LHS QPO spectrum as the amount of
reflection changes with QPO phase. 

By contrast, in the SIMS, the extent of the flow is so small ($r_i=7$
and $r_o=10$) that even this large misalignment angle does not give
rise to significant variability in the total reflection fraction.
Previous rms spectral analysis of the QPO have not looked at this in
detail (e.g. Sobolewska \& Zycki 2006). We plan to address this issue
in a future work (Axelsson et al in preparation). 

\section{Conclusions}
\label{sec:conc}

The truncated disc / precessing inner flow model for the spectral timing properties
of XRBs predicts a QPO phase dependence of the iron line profile. 
This results from the inner flow preferentially illuminating different
regions of the disc as it precesses. When the brightest region of the disc is moving
towards us, the iron line will be blue shifted and boosted. When the brightest
region is receding, the iron line will be red shifted. As the illumination pattern
rotates around the disc, the iron line rocks between blue and red shift.
This process always happens in a particular order with the most heavily blue shifted
iron line profile following the QPO trough and the most heavily red shifted
iron line profile following the QPO peak. It is possible to isolate the peaks
and troughs in a light curve using a simple flux selection. The rising phase, which
follows the trough, is predicted to have the bluest iron line and the falling phase,
which follows the peak, is predicted to have the reddest iron line.

We predict this QPO phase dependence of the iron line profile to be present for a
large range of spectral states (and therefore truncation radii). This means that
it may be best to search for the effect in spectra containing type-B QPOs which
have very little broad band variability associated with them and therefore provide a
much cleaner signal than type-C QPOs. However, the nature of the iron line phase
dependence changes with truncation radius. When it is large,
the red wing can dominate over the blue wing during the fall from QPO peak to trough.
When it is small, Doppler and relativistic boosting from the rapidly moving inner
regions of the disc
means that the red wing can never dominate over the blue wing. The characteristic
shape of the difference spectrum between rise and fall should therefore change as the
spectrum evolves from the LHS to the SIMS. The dominant red wing of the QPO fall
spectrum in the LHS (the `red dip' in the difference spectrum) is the most unique
model prediction but if we wish to observe this, we must disentangle the underlying
QPO signal from the broad band noise. This will be the subject of a future paper. An
observation of the effect in both states, along with confirmation that the difference
spectrum \textit{changes} with state, would constitute excellent evidence, not only
of the precession model, but also that the truncation radius moves between the LHS and
the SIMS.

Quasi-periodic shifting of the iron line peak energy is a unique
prediction of the Lense-Thirring precession model for the low
frequency QPO in XRBs. We have shown that it may be possible to
observe such an effect with current missions, but that \textit{LOFT}
will be able to  measure this with precision, enabling us to place
accurate constraints on the accretion geometry.

\section{Acknowledgements}

AI acknowledges the support of an STFC studentship. AI thanks Sergio Campana for
useful discussions and also work to measure the properties of the LHS and SIMS
models. AI also thanks Luigi Stella and Micheil van der Klis for many useful
discussions.

\appendix

\section{Geometry}
\label{sec:geomcalc}

In order to perform our calculations, we must define some vectors using the coordinate
system outlined in Figure \ref{fig:schem}. We represent the $x$, $y$ and $z$ axes with
the standard $\hat{\underline{i}}$, $\hat{\underline{j}}$ and $\hat{\underline{k}}$
unit basis vectors. It then follows from Figure \ref{fig:schem} that
\begin{eqnarray}
\hat{\underline{J}}_{BS} &=& -\sin{\beta}~\hat{\underline{j}}
                             +\cos{\beta}~\hat{\underline{k}} \nonumber \\
\hat{\underline{\epsilon}} &=& \cos{\beta}~\hat{\underline{j}}
                              +\sin{\beta}~\hat{\underline{k}}.
\end{eqnarray}
The three vectors $\hat{\underline{i}}$, $\hat{\underline{\epsilon}}$ and
$\hat{\underline{J}}_{BS}$ therefore form a right handed Cartesian coordinate system:
the disc basis vectors. We can define a vector, $r_d~\hat{\underline{r}}_{d}$, which
points from the origin (the black hole) to any point on the disc where
\begin{equation}
\hat{\underline{r}}_{d} = \cos{\phi_d} ~\hat{\underline{i}}
                            +\sin{\phi_d} ~\hat{\underline{\epsilon}}.
\end{equation}
Note, because the disc is razor thin, there is no $\hat{\underline{J}}_{BS}$ component
(i.e. $\hat{\underline{J}}_{BS}.\hat{\underline{r}}_{d}=0$) and $\phi_d$ is simply the
angle between $\hat{\underline{r}}_{d}$ and the x-axis. We also define a vector
pointing from the origin to the observer using the disc basis vectors
\begin{equation}
\hat{\underline{S}} = \sin{\theta_i}\cos{\phi_i}~\hat{\underline{i}}
                     +\sin{\theta_i}\sin{\phi_i}~\hat{\underline{\epsilon}}
                     +\cos{\theta_i}~\hat{\underline{J}}_{BS}.
\end{equation}

In order to describe points on the surface of the flow, we must define flow
basis vectors. The `z-axis' of this right handed coordinate system is
$\hat{\underline{J}}_{flow}$ which precesses around $\hat{\underline{k}}$ as
illustrated in Figure \ref{fig:schem}. The other two basis vectors,
$\hat{\underline{x}}_{f}$ and $\hat{\underline{y}}_{f}$, must therefore
also precess with the flow. We use
\begin{eqnarray}
\hat{\underline{x}}_{f} &=& \cos{\gamma}~\hat{\underline{i}}
                           +\sin{\gamma} ~\hat{\underline{j}}
 \nonumber \\
\hat{\underline{y}}_{f} &=& -\cos{\beta}\sin{\gamma}~\hat{\underline{i}}
                            +\cos{\beta}\cos{\gamma}~\hat{\underline{j}}
                            +\sin{\beta}~\hat{\underline{k}}
 \nonumber \\
\hat{\underline{J}}_{flow} &=& \sin{\beta}\sin{\gamma}~\hat{\underline{i}}
                              -\sin{\beta}\cos{\gamma}~\hat{\underline{j}}
                              +\cos{\beta}~\hat{\underline{k}},
\end{eqnarray}
such that $\hat{\underline{x}}_{f} = \hat{\underline{i}}$ when $\gamma=0$ but, as the
precession angle unwinds, the axes move. We can then specify a point in the flow
with the vector $r_f~\hat{\underline{r}}_{f}$ where
\begin{equation}
\hat{\underline{r}}_f = \sin{\theta_f}\cos{\phi_f}~\hat{\underline{x}}_f
                       +\sin{\theta_f}\sin{\phi_f}~\hat{\underline{y}}_f
                       +\cos{\theta_f}~\hat{\underline{J}}_{flow}.
\end{equation}
Here, $\theta_f$ is the angle between $\hat{\underline{r}}_f$ and
$\hat{\underline{J}}_{flow}$ and $\phi_f$ is the angle between
$\hat{\underline{r}}_f(\theta_f=\pi/2)$ and $\hat{\underline{i}}$.

Because our flow is elliptical with semi-minor axis in the
$\hat{\underline{J}}_{flow}$ direction and semi-major axis in the
$\hat{\underline{a}} = \cos{\phi_f}~\hat{\underline{x}}_f
+\sin{\phi_f}~\hat{\underline{y}}_f$ direction, the distance from the origin to
any point on the surface is
\begin{equation}
r_f(\theta_f) = \frac{r_o h_o}
              {\sqrt{(h_o \sin{\theta_f})^2 + (r_o \cos{\theta_f})^2}}.
\end{equation}
Because $r_f$ is uniquely determined by $\theta_f$, we can define
$dr = |r(\theta_f)-r(\theta_f+d\theta_f)|$.

We need to be able to write down the unit vector normal to the flow surface. We
can do this using a few identities. Imagine a triangle drawn between the
two focuses of the ellipse, $F_1$ and $F_2$, and any point on the circumference
of the ellipse, $P$. We know that the distance from the origin to either focus
is $f=\sqrt{r_o^2-h_o^2}$ and also that the three sides of the triangle add up
to $2r_o+2f$. We can define the angle between the line from $P$ to $F_1$ ($P~F_1$)
and the line from $P$ to $F_2$ ($P~F_2$) as $\psi$. We know that the surface area
unit vector, $\hat{\underline{A}}$, goes directly between these two lines such
that the angle between $-\hat{\underline{A}}$ and each line is $\psi/2$. We can
say that $\hat{\underline{A}}$ points from some point $x_o~\hat{\underline{a}}$
to the point on the flow surface, $P$, in such a way that this condition is
satisfied. Say that $d$ is the distance from $P$ to $F_2$ and $\Omega$ is the
angle between the lines $F_2~F_1$ and $F_2~P$. We can use the cosine rule
a few times to show that $d=\sqrt{f^2+r_f^2-2fr_f\sin{\theta_f}}$ and
$\cos{\Omega}= (f^2-r_o^2+r_od)/(f~d)$. It is then possible to show that
\begin{equation}
\hat{\underline{A}} = \frac{r_f~\hat{\underline{r}}_f - x_o~\hat{\underline{a}}}
                      {\sqrt{x_o^2+r_f^2-2x_or_f\sin{\theta_f}}}
\end{equation}
where 
\begin{equation}
\cos{\psi} = \frac{2r_o^2+d^2-2r_od-2f^2}{d(2r_o-d)}
\end{equation}
and
\begin{equation}
x_o = f - \frac{d\sin(\psi/2)}{\sin(\pi-\psi/2-\Omega)}.
\end{equation}

We will also need to define a vector which points from a given point on the
flow to a given point on this disc. This can be written as
\begin{equation}
\zeta~\hat{\underline{\zeta}} = -r_f~\hat{\underline{r}}_f
                                +r_d~\hat{\underline{r}}_d.
\end{equation}
From this, it is simple to show that the distance between the two points is
\begin{equation}
\zeta^2 = r_f^2+r_d^2-r_fr_d~\hat{\underline{r}}_f . r_d~\hat{\underline{r}}_d.
\end{equation}
All of these vectors will become very useful for the following sections.

\section{Disc irradiation calculations}
\label{sec:disccalc}

So, we need to calculate what luminosity a disc element with surface area
$dA_d=r_d d\phi_d dr_d$ will intercept from a flow surface element emitting a
luminosity $dL$ over a semi-sphere (because it only emits \textit{away} from the
rest of the flow). We can then integrate over all flow elements to work out the
total flow luminosity that the disc element intercepts. For the disc patch to
see anything at all from a given flow element, it must pass two tests. First,
does it lie in the unit semi-sphere of the flow element; i.e. is
$\hat{\underline{A}} . \hat{\underline{\zeta}} > 0$. Also, because we are
viewing the top of the system ($\theta_i \leq 90^o$), we only see luminosity
which has reflected off the \textit{top} of the disc. Therefore, we only count
luminosity incident on the top of the disc in our integral. This means we require
$\hat{\underline{\zeta}} . \hat{\underline{J}}_{BS} < 0$. If one of these
conditions isn't met, the luminosity intercepted by the disc element is
$dL_r=0$. If both are, we have
\begin{equation}
dL_r = \frac{(-\hat{\underline{\zeta}} . \hat{\underline{J}}_{BS})
dA_d}{2\pi\zeta^2} dL.
\end{equation}
We see that, the amount of luminosity intercepted depends on the projected area
of the disc patch as seen by the flow element. If the patch is face-on as seen
by the flow, $\hat{\underline{\zeta}} . \hat{\underline{J}}_{BS}=1$ and the
projected area is $dA_d$. This area reduces as the patch turns away from the
emitting flow element. The total luminosity incident on a disc patch is
calculated by adding up the contribution from every flow element.

\section{Flow modulation calculations}
\label{sec:flowcalc}

We now need to calculate how much luminosity a telescope with effective area
$A_{eff}$ will intercept from a given flow element in order to again integrate
over the whole flow. For the telescope to see any luminosity at all, two tests
must again be passed. First of all, the viewer must be in the unit semi-sphere
of the flow element. This means we require
$\hat{\underline{A}} . \hat{\underline{S}} > 0$. We also wont see anything
if the emission is blocked by the disc. We know the emission definitely wont be
blocked by the disc if the flow element is above the disc; i.e.
$\hat{\underline{r}}_f . \hat{\underline{J}}_{BS} > 0$. Even if the element
is below the disc plane, we still might be able to see through the hole in
the centre of the disc. So, imagine a point on the flow which is below the
disc plane, emitting along the vector $\hat{\underline{S}}$. At some point it
will intercept the disc plane. The distance between the flow element and the
point where the vector crosses the disc plane is $\zeta$. This point will be a
distance $r_d$ from the origin. We can write
\begin{equation}
\zeta \hat{\underline{S}} = -r_f~\hat{\underline{r}}_f
                            +r_d~\hat{\underline{r}}_d.
\end{equation}
Dotting both sides with $\hat{\underline{J}}_{BS}$ and rearranging gives
\begin{equation}
\zeta= \frac{-r_f~\hat{\underline{r}}_f . \hat{\underline{J}}_{BS}}
            {\cos{\theta_i}}.
\end{equation}
We then know that
\begin{equation}
r_d^2 = \zeta^2 + r_f^2 + 2\zeta r_f \hat{\underline{S}}
 . \hat{\underline{r}}_f.
\end{equation}
If $r_d^2 < r_o^2$, we still see the flow element through the hole in the disc.
If not, it is hidden by the disc.

So, if the unit-sphere and disc obstruction tests are not passed, the luminosity
intercepted by the telescope is $dL_{obs}=0$. Otherwise, this is
\begin{equation}
dL_{obs} = \frac{A_{eff}~dL}{2\pi D^2},
\end{equation}
where $D$ is the distance to the source. Note, because the telescope is so far
away and is pointed straight at the black hole, we can say that the projected
area of the telescope as seen by any flow element is $A_{eff}$. We then just
set $A_{eff}/(2\pi D^2)=1$, because it only tells us about normalisation, and
sum up the contribution from each flow element.

\section{Iron line profile calculations}
\label{sec:ironcalc}

A disc element at $r_d~\hat{\underline{r}}_d$ is rotating with Keplerian velocity
$v_k$. An observer at $\theta_i$, $\phi_i$ then sees the disc patch travelling towards
them at a velocity of $v=v_k \sin{\phi} \sin{\theta_i}$ where $\phi=\phi_i-\phi_d$.
The tangent points of the disc will therefore travel towards the observer
at a velocity of $\pm v_k \sin{\theta_i}$. This means that a photon emitted with
energy $E_{em}$ with be red shifted by
\begin{equation}
E_{em}/E_{obs}=(1-3/r_d)^{-1/2} 
\left[   1 + \frac{\cos{\alpha}} { [ r_d(1+\tan^2{\xi_o})-2 ]^{1/2}   }  \right],
\end{equation}
where
\begin{eqnarray}
\cos{\alpha} &=&\sin{\phi}\sin{\theta_i}(\cos^2{\theta_i}
               +\cos^2{\phi}\sin^2{\theta_i})^{-1/2}
\nonumber \\
\tan{\xi_o} &=& \cos{\phi}\sin{\theta_i}(1-\cos^2{\phi}\sin^2{\theta_i})^{-1/2},
\end{eqnarray}
(Fabian et al 1989; 2000).

For a given precession angle, $\gamma$, the flow luminosity incident on a
disc patch described by $r_d$ and $\phi_d$ is $L_r(r_d,\phi_d)$. If this
luminosity were all emitted at energy $E_{em}$, the observer would see a
luminosity, all at $E_{obs}$, of
\begin{equation}
dL_{obs} \approx L_r(r_d,\phi_d) (E_{obs}/E_{em})^3 \cos{\theta_i}.
\end{equation}
Here, the approximations come from assuming light to travel in a straight
line. Throughout this paper, we ignore gravitational light bending thus
taking these to be good approximations. This should be apropriate since
the inner radius of the flow is assumed to be $r_i=7$ throughout and light
bending effects outside of this radius will be minimal. The total observed
luminosity as a function of energy is calculated by summing the contribution
from each disc patch. As the flow precesses and the function 
$L_r(r_d,\phi_d)$ evolves, the observed iron line profile will change.

\label{lastpage}

\end{document}